\documentclass[journal]{IEEEtran}
\usepackage{xcolor,soul,framed} 
\usepackage{verbatim}
\usepackage{caption}
\usepackage{subcaption}
\usepackage[utf8]{inputenc}
\usepackage[ruled,vlined]{algorithm2e}
\usepackage{algpseudocode}
\colorlet{shadecolor}{yellow}
\usepackage{cite}
\usepackage[pdftex]{graphicx}
\graphicspath{{../pdf/}{../jpeg/}}
\DeclareGraphicsExtensions{.pdf,.jpeg,.png}

\usepackage[cmex10]{amsmath}
\usepackage{array}
\usepackage{mdwmath}
\usepackage{mdwtab}
\usepackage{eqparbox}
\usepackage{url}

\begin{document}
\bstctlcite{IEEEexample:BSTcontrol}
    \title{ Mesh of Things (MoT) Network-Driven Anomaly Detection in Connected Objects}
  \author{Rathinamala Vijay, 
      Prabhakar. T. V.,\\
   Department of Electronic Systems Engineering, \\
   IISc,  Bangalore, India
}


\maketitle

\begin{abstract}
This paper presents a hybrid Mesh of Things (MoT) network performance model to evaluate the end-to-end Packet Delivery Ratio (PDR) and latency.  These PDR and latency measures are used to identify both a de-tangled mesh as well as to track the mesh successfully. A de-tangled mesh is a mesh with an anomaly where one or more nodes are separated from the rest of the mesh network. We demonstrate the performance model of a hybrid BLE mesh-PLC network by considering an air cargo monitoring application and validate with experimental PDR, and latency data. The link uncertainty in Bluetooth Low Energy (BLE) mesh may be attributed to (a) RF interference,  (b)~Transmitter's vicinity range, and (c) Receiver sensitivity. In contrast, the link uncertainty in Power Line Communication (PLC) may be attributed to: (a) Colored background noise, (b)~Channel frequency response, and (c) Impulse noise appearing due to load state as well as variations in the powerline. In our work, we construct an equivalent Bayesian network for the mesh to be tracked, capture the uncertainty within the mesh links using the Noisy-OR and the Noisy-Integer addition model and perform belief propagation to detect and localize a network anomaly.
\end{abstract}

\begin{IEEEkeywords}
Anomaly detection, Latency, PDR, Bayesian network, Generalized Noisy-OR, Belief propagation, shortest path model, BLE, PLC, Performance model
\end{IEEEkeywords}

%
\IEEEpeerreviewmaketitle


\section{Introduction}
\begin{figure} [!h]
    \centering
    \includegraphics[width=\columnwidth]{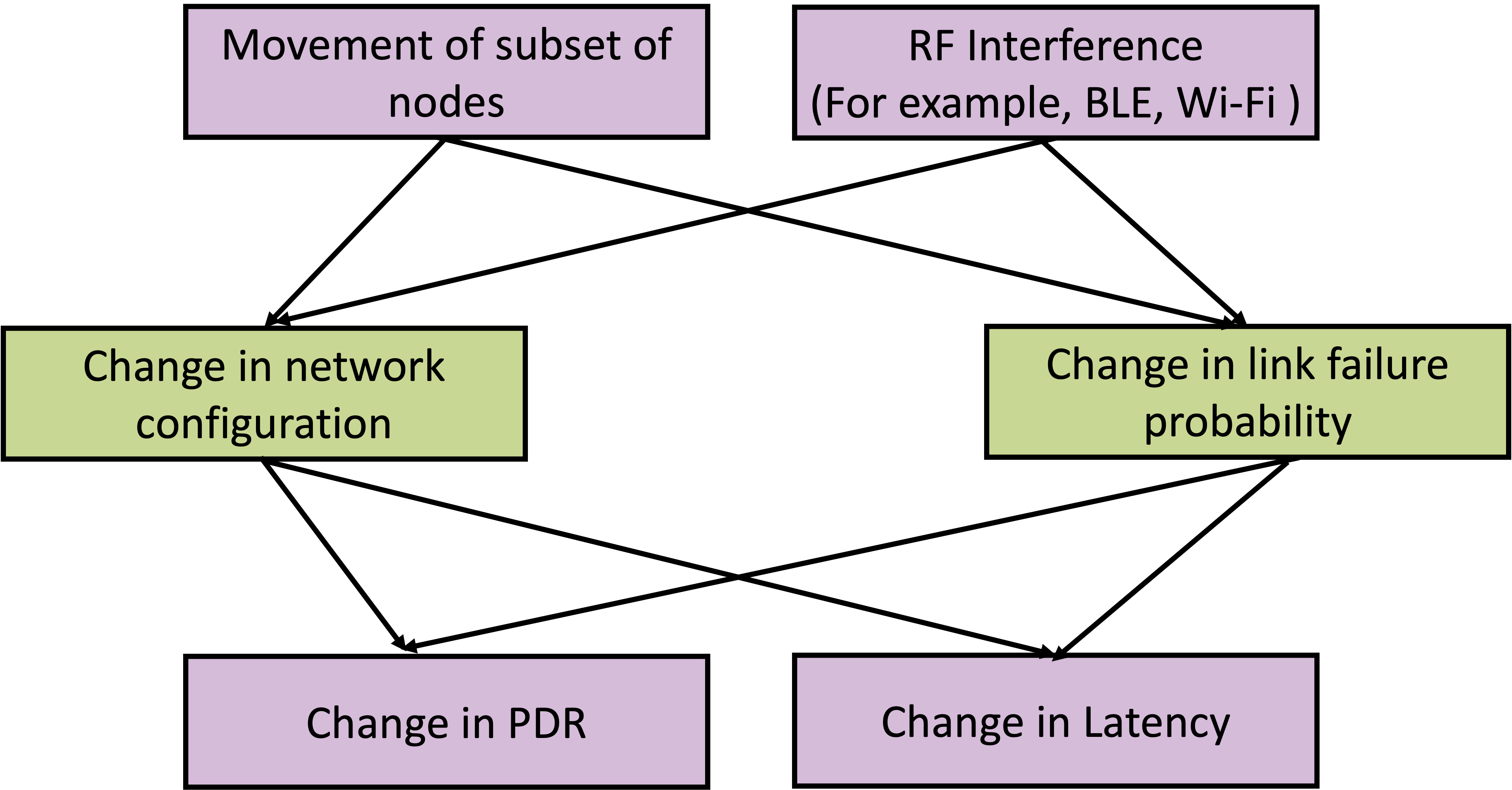}
    \caption{Monitoring MoT network with measures such as latency and Packet Delivery Ratio (PDR). Movement of nodes and RF interference impacts these measures, and when there is a significant change, it will help in identifying and reporting a de-tangled mesh.}
    \label{fig:MoT_char}
\end{figure}

A Mesh of Things (MoT) network connects multiple objects using redundant paths and overcomes the case of single-point failure. This network is  useful in monitoring a group of objects in many applications. Applications of the MoT network include medical equipment tracking, farm animal tracking, freight dispatch from the warehouse, restricted area access monitoring, etc. Characterizing and monitoring such a network plays a crucial role in increasing the efficiency of tracking applications.\\
Latency and PDR are two measures that are useful in characterizing an MoT network. Fig.~\ref{fig:MoT_char} portrays the impact of two factors, namely, \textit{Movement of nodes} and \textit{RF interference} on Latency and PDR measures. The movement of a subset of nodes away from the mesh in an MoT network will not only introduce a change in the mesh configuration but will also impact the end-to-end latency and PDR. Additionally, RF interference from nearby sources such as BLE, Wi-Fi, and other technologies affects the MoT communication and the associated latency and PDR. In our work, we define a ``de-tangled mesh" in the backdrop of two scenarios: (a) When the inter-node distance between the mesh nodes increases from the dense baseline, basically forming a sparse network, or (b)~When one or more nodes breaks away or disconnects from the mesh. Thus, the latency and PDR  measures characterize an MoT network. The baseline and a de-tangled mesh network can be easily classified using this characterization. \\
\\
Continuous monitoring of the performance measures aids in real-time tracking of every object in the mesh network efficiently. The object that is disconnected  from the network can be identified when comparing the current measures of MoT with that of the measures of the characterized baseline and de-tangled states of mesh. \\ 

 We design and implement a heterogeneous MoT network testbed for demonstrating a smart cargo monitoring application and realize this network using BLE and PLC technologies. Such a heterogeneous network seamlessly extends the range of connected objects with minimum latency overheads to accomplish smart cargo monitoring. We describe the characterization of this MoT network through performance measures and perform real-time monitoring of a group of things in the MoT network.

 \begin{figure} [htbp]
     \centering
     \includegraphics[width=\columnwidth]{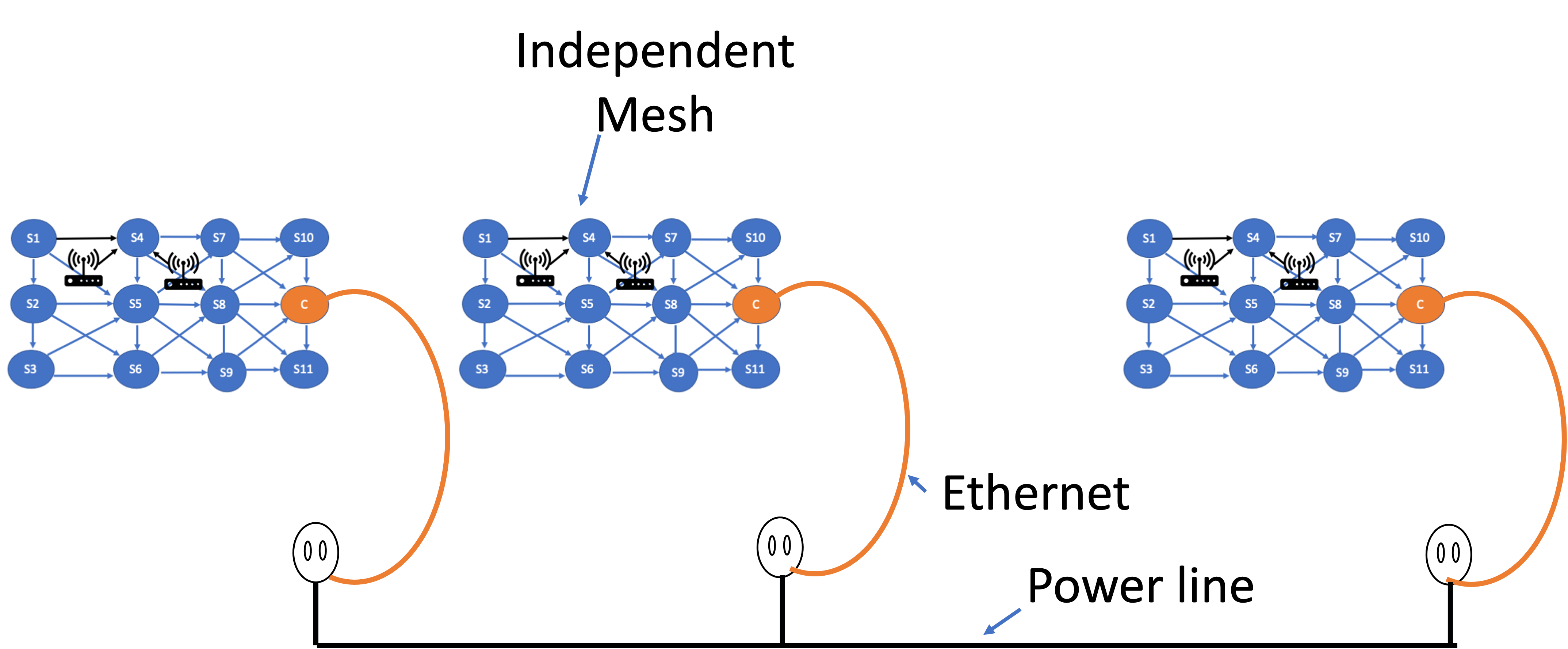}
     \caption{Hybrid BLE-PLC network to cover a large area as an efficient and scalable alternative to the large BLE mesh network. The mesh near any power outlet can be an entry point to the Power Line Communication.}
     \label{fig:lpwan_ch5}
 \end{figure}
 
\noindent \textit{Contributions:}
\begin{itemize}
    \item We construct a generalized apriori Bayesian network model to detect a de-tangled MoT network using PDR measures. The average latency within a mesh is numerically computed using Dijkstra's shortest path algorithm.
    \item We analyze the scalability of the MoT network by extending the mesh range with a backbone PLC network. We characterize the average PDR and latency of the PLC  link from its channel frequency response model.
    \item We conduct extensive numerical simulations to evaluate the average PDR and latency under several possible scenarios, such as vicinity range, RF interference, and impulse noise. We validate these measures using our experimental results. 
\end{itemize}
The rest of the paper is organized as follows. We describe the related work in section~\ref{sec:related_work}. We introduce the characterization of the BLE mesh network in section~\ref{sec:ble_mesh_char}. In section~\ref{sec:reliability}, we describe the reliability analysis for the BLE mesh network. In section~\ref{sec:mesh_lat}, we describe the latency analysis for the BLE mesh network. Further, we describe the method for anomaly  detection in mesh through the performance measures in section~\ref{sec:detangled}. Furthermore, we introduce the characterization of the PLC network in section~\ref{sec:PLC_char}. In sections~\ref{sec:plc_lat} and \ref{sec:plc_rel}, we describe the latency and reliability analysis of the PLC network, respectively. In sections~\ref{sec:hybrid_rel} and \ref{sec:hybrid_lat}, we formulate the reliability and latency expressions in a hybrid MoT network. In section~\ref{sec:disc}, we answer a few research questions and present some insights from this work.
\section{Related Work}
\label{sec:related_work}
Although the BLE mesh network is widely deployed to realize high reliability and low latency applications, the scalability of this network remains a challenge \cite{rondon2019understanding,baert2018bluetooth,ghori2020bluetooth,BLE_scale_issue,BLE_scale_issue1, dense_mesh_issue, hernandez2020bluetooth,conn_char}. 
The flooding mechanism in a large BLE mesh protocol increases the overall traffic in the mesh and increases collisions in the network. The multi-packet collisions induce re-transmissions of the source nodes. Thus, a large mesh network deployment has scalability issues. The literature suggests several mitigation methods such as (a) Random back-off before a packet transmission \cite{darroudi2020experimental,baert2018bluetooth,rondon2017evaluating}, (b)~Inter-packet randomization \cite{rondon2019understanding}, (c) relay selection techniques \cite{Reno_relay_selection}, and (d) isolated data and control plane usage with hybrid BLE Mesh \cite{murillo2020all}. In our work, we overcome the
need for building a large mesh and the associated complex countermeasures to reduce collision. \\

In literature, there are works that propose a hybrid network to implement wide area coverage \cite{Lora_Ble1, Lora_Ble2, Zig_Ble, WiFi_Ble}; the authors of \cite{leonardi2022lora} design the LoRaBLE network to connect independent clusters of BLE with LoRa and evaluate its performance. The authors of \cite{Lora_Ble} use the hybrid LoRa and BLE technologies to monitor wildlife over a wide area and present the performance of such a network. We propose to restrict the size of the mesh and use it in combination with PLC devices to extend the range. In this way, multiple independent dense meshes can be connected using the backbone powerline network.  In general, the PLC network suffers from colored background noise and impulse noise due to load variations. However, these PLC devices have the ability to do channel estimation for the best channel selection before transmission and hence have a better wide area coverage with fewer PLC nodes. The combination of BLE and PLC networks  \cite{mala2019heterogeneous} are efficient in implementing a low power wide area network that needs lesser focus on scalability issues. Fig.~\ref{fig:lpwan_ch5} shows an example hybrid BLE-PLC network where several independent meshes can be interconnected using a backbone PLC network. This scenario applies to many applications, such as smart industrial sensors distributed across different building floors or medical equipment spread across floors or in adjacent buildings. \\

In this work, we detect anomalies in a mesh network using PDR and latency computed using Bayesian belief networks \cite{baynet1, sethi2004, kandula2005shrink, bahl2007towards, liu2008passive} and the shortest path model \cite{dijkstra1959note}, respectively. The uncertainty in the performance metrics introduced by RF interference is modeled using Noisy-OR and the Noisy integer addition function integrated with the Bayesian network \cite{srinivas1993generalization}.  Most of the prior work in the literature detects network anomalies in an enterprise network with a fixed topology. In contrast, our work detects a network anomaly in any application where a group of connected objects is required to stay together or move together to a new destination.
\section{BLE mesh network characterization}
\label{sec:ble_mesh_char}
\begin{figure}[!h]
    \centering
    \includegraphics[width=0.8\columnwidth]{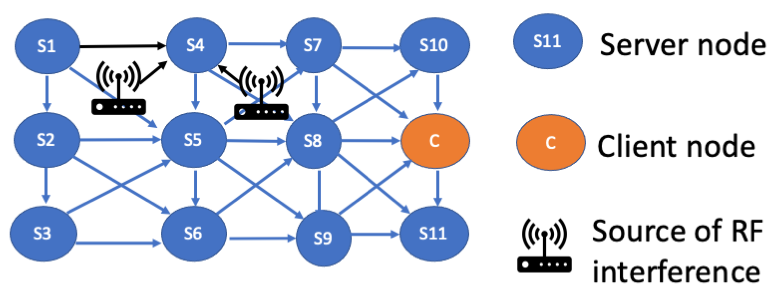}
    \caption{BLE mesh with server and client nodes with RF interference sources. The distance between each neighboring server node is 0.6 m. }
    \label{fig:RF_inter_ch5}
\end{figure}

Our goal is the detection of a de-tangled mesh and as well as continuous tracking of a mesh using performance measures such as PDR and latency.  We first describe the simulation setting to demonstrate the tracking of performance measures in a baseline and a de-tangled mesh.\\
\noindent \textbf{Simulation setting:}\\
Fig.~\ref{fig:RF_inter_ch5} shows the baseline BLE mesh where a client node C, aggregates data from all server nodes S1-S11. 
 The client node selects a subset of server nodes within the mesh  and sends a request to them. This aggregation method reduces the probability of collision in a mesh compared to the method of collecting data from all server nodes simultaneously. Selection of nodes within the mesh network \cite{hansen2018relay} can be broadly classified as energy-based or geography-based selection methods, but we choose a simple odd or even numbered node selection method. All nodes considered here are more or less similar geographical positions, and the energy associated with each node is identical; our simple method suffices for the choice of partitioning the mesh nodes and collecting data from them. Our case study considers a mesh with eleven server nodes and a client. 
Each individual group of nodes acts as a source node sending a response packet to a client's request, while all other nodes in the mesh act as relay nodes. Here the client is the destination node. Every source node typically takes 2-hop, 3-hop, 4-hop, or i-hop to deliver a packet to the destination node. We begin by characterizing PDR and latency measures for a baseline setup and subsequently track these measures.\\
\textbf{Assumptions:}\\
We perform the numerical simulations of the PDR  and latency measures in an MoT network under the following assumptions:
\begin{itemize}
    \item In our baseline mesh, we consider that the BLE mesh server and client nodes always move together.
    \item All nodes in the mesh transmit with the same power based on the scenario considered. For example, in Scenario 1, all nodes transmit with 4 dBm power.
    \item If one or more nodes get separated from the mesh; it is considered a de-tangled mesh.
    \item The average RF interference, colored background noise, PDR, and latency across the mesh units are considered independent and identically distributed (i.i.d). 
    \item The client-to-gateway communication is considered to be reliable with an ethernet connection to a PLC modem. The simulations do not include the processing of ethernet packets.
    \item The gateway has the capability to encapsulate the BLE packet to PLC and vice versa.
\end{itemize}
\section{BLE mesh reliability analysis}
\label{sec:reliability}
\begin{figure}[htbp]
     \centering
     \includegraphics[width=0.8\columnwidth]{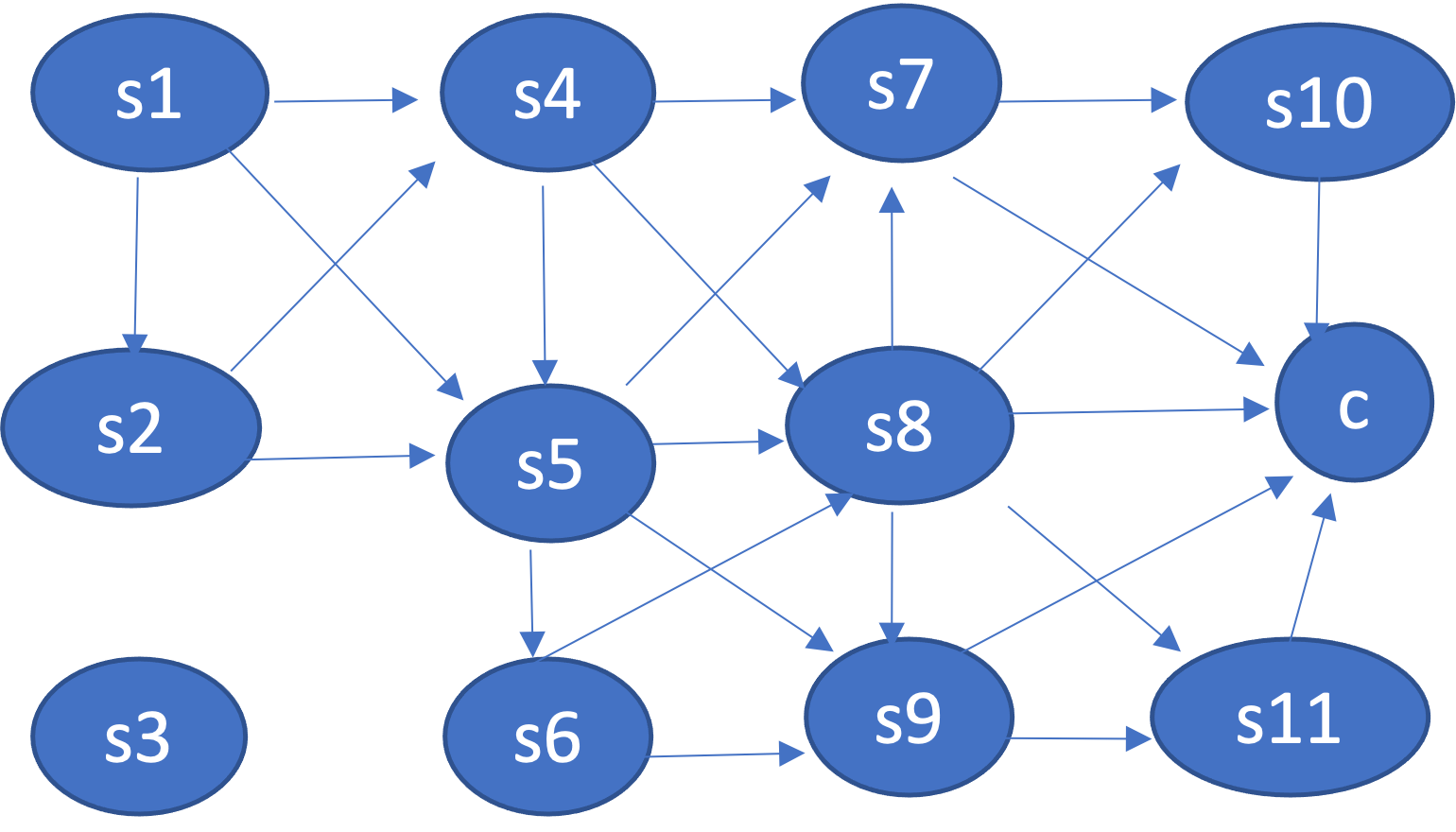}
     \caption{Bayesian network equivalent of an MoT network with binary state nodes and Noisy-OR function at multiple parent nodes. All possible 4-hop paths between the server node S1 to the client node C are shown.}
     \label{fig:bin_state_nodes}
 \end{figure}

   To model PDR for a mesh network, we group all possible i-hop paths between the source-destination pair and construct an equivalent a priori Bayesian network. For example,   we combine all possible 4-hop paths between a source node and client node in a baseline mesh shown in Fig.~\ref{fig:bin_state_nodes}. Next, we use the belief propagation algorithm \cite{baynet1} and conduct Bayesian inference on a priori network to calculate the belief for a packet to be received at the destination given a packet is transmitted at the source. 
   This process is repeated for each source-destination pair among every group, and an average PDR is calculated within each group. The overall average PDR for the entire mesh network is the average PDR across the two groups. \\
   The link failure probabilities of each link in this mesh network take values uniformly in the interval [0,1].  We simulate 1000 independent link failure probability states indicating the time-varying nature of the links in the mesh. This link failure probability captures the impact of RF interference, receiver sensitivity, and transmit level of each node in the mesh network.\\
\subsection{Bayesian network with binary state nodes and Noisy-OR function}
\begin{figure}
    \centering
    \includegraphics[width=\columnwidth]{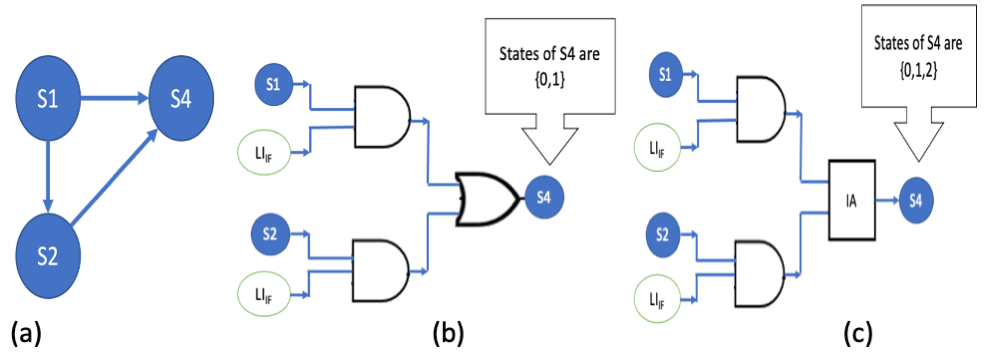}
    \caption{(a) Shows a portion of the Bayesian network, (b) and (c) show the basic operation of Noisy-OR and Noisy-Integer addition function and their corresponding states of the node S4. $L1_{IF}$ and $L2_{IF}$ are the link interference associated with nodes S1 and S2.}
    \label{fig:nor_ia}
\end{figure}
Fig.~\ref{fig:bin_state_nodes} shows a mesh network where the adjacent nodes are placed 0.6 m away from each other. 
The power level chosen allows the packet from a source node to take two or more hops to reach the destination node. We collect the set of 2-hop, 3-hop, or i-hop paths separately. 
We construct an equivalent Bayesian network for the given mesh network for each i-hop group separately. We perform belief propagation whenever there is a new packet generated in the mesh. The failure probability for each link in the mesh depends on RF interference, transmitter vicinity range, and receiver sensitivity. 
Our numerical simulation results indicate that the PDR for 4-hop, 3-hop, and 2-hop paths between source and client averaged over 1000 independent link failure scenarios yields 82.3\%, 91.9\%, and 99.5\%, respectively. Now under a realistic scenario, there can be combinations of i-hop paths. Hence, we use different ratios for each hop group to realize scenarios emulated for different transmit node powers and interference levels.\\
\begin{equation}
PDR_{ble}=\sum_{i=0}^{N}k_i \cdot PDR_i\\
\label{eq:mesh_pdr}
\end{equation}
where $PDR_{ble}$: is the average PDR in the considered mesh network.\\
$k_i$: a fraction of i-hop paths between the client and server nodes.\\
$i$: number of hops between the client and server nodes. \\
$PDR_i$: average PDR of the i-hop paths between the client and server nodes.\\
\subsection{Bayesian network with discrete state nodes and Noisy-Integer addition function}
\begin{figure}
    \centering
    \includegraphics[width=0.8\columnwidth]{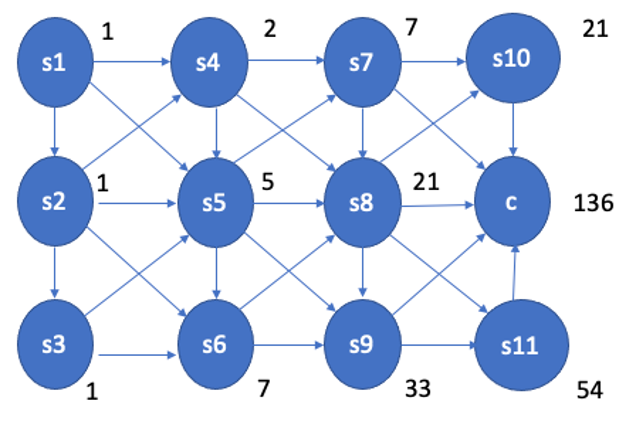}
    \caption{Bayesian network equivalent to experimental MoT network, with discrete state nodes and integer addition function at each node. This network is used to evaluate the belief for the number of paths that exist between the source and the destination nodes. For example, the number of paths from the source node S1 to various destination nodes is represented at the right of each node in this figure. The existence of multiple paths ensures high reliability in the network.}
    \label{fig:int_add}
\end{figure}

Fig.~\ref{fig:int_add} is an equivalent Bayesian network for a mesh network shown in Fig.~\ref{fig:RF_inter_ch5}, where each node is a random variable that take discrete state values. The state of each node depends on the state of its parents and is computed using a Noisy-integer addition function.  This network takes into account all possible hop paths for a node from the source node in a single shot. 

 The algorithm, as in \cite{srinivas1993generalization}, is applied to estimate the state of each node. The node's state represents the number of redundant paths to the node from the source node. This quantifies the reliability of the link between a pair of nodes. The conditional probability of the state of a node given the state of the parents and link failure probabilities between the node and each of its parents are used to evaluate the network reliability. Hence, the reliability model helps characterize an MoT network.
\section{BLE mesh latency analysis}
\label{sec:mesh_lat}
Dijkstra's shortest path algorithm \cite{dijkstra1959note} is used to determine the minimum latency between a server node and a client node. 
The transmit power of all server nodes is set to different levels to achieve various vicinity ranges. The vicinity range is inversely proportional to the number of hops to reach the client node. 
 We consider three scenarios for our latency simulations. The mesh latency evaluation is carried out with three different power levels in each scenario.\\
 \begin{figure}[!h]
    \centering
    \includegraphics[width=0.6\columnwidth]{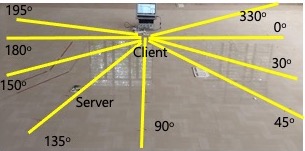}
    \caption{Experimental setup  showing a BLE mesh client node connected to a computer through the serial port to issue commands. The server node placed at various distances and angles responds to the client's queries.} 
    \label{fig:vic_measure}
\end{figure}  
\begin{figure}[!h]
    \centering
    \includegraphics[width=0.6\columnwidth]{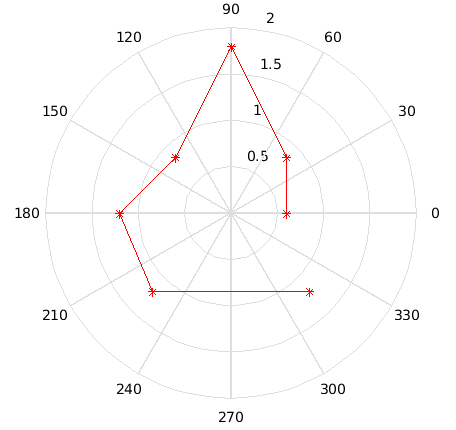}
    \caption{Experimental setup for vicinity range with transmit power set to -4 dBm. Receiver distance and angle are varied during measurements. The receiver distance range is 0.6 m - 3 m, and the receiver angle ranges from 0 to 360 degrees with respect to the transmit node.}
    \label{fig:vr_pattern}
\end{figure}

\begin{figure}[h!]
    \centering
    \includegraphics[width=\columnwidth]{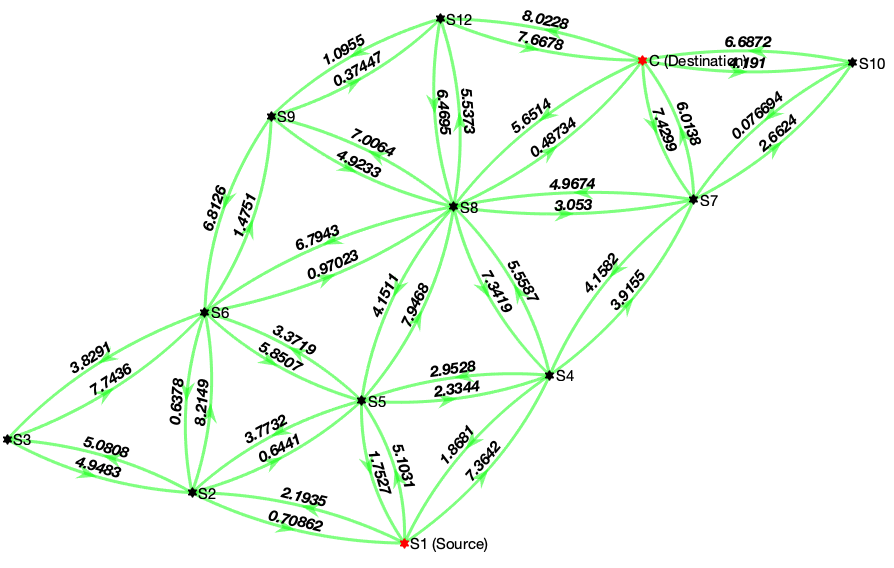}
    \caption{Scenario 1: The neighbor nodes have a stable connection. The vicinity range of each node considered in the simulation is 0.86 m. The average source-to-destination latency obtained in this scenario is 6.23 ms, while experimental results show 6.30 ms.}
    \label{fig:scene_4dB}
\end{figure}

\noindent\textbf{Experimental setup for vicinity range measurements}\\
 Vicinity range measurements between a client node and a server node help in mesh network modeling. 
We place the mesh client node fixed at a location while the server node position varies. The position varies in distance ranging from 0.6 m - 3 m, and the orientation varies  from 0 to 360 degrees. The experimental setup is shown in Fig.~\ref{fig:vic_measure}. The server node is set at a power of -4 dBm. The coverage distance of the source-to-destination pattern is shown in Fig.~\ref{fig:vr_pattern}. We embed the vicinity range deduced through measurements for each node correspondingly into the simulation. The destination node's orientation and distance with respect to the source node decide the coverage range. This experiment helps in the placement of each node at appropriate angles to achieve maximum coverage. 
\subsection{Scenario 1: Vicinity range (0.86 m) with 4 dBm power}
Scenario 1,  shown in Fig.~\ref{fig:scene_4dB}, has a 12-node network with a horizontal and vertical distance between each node is 0.6 m while the diagonal distance is 0.86 m. Scenario 1 has all nodes transmitting with a maximum power of 4 dBm using an NRF52832 board. The Vicinity range achieved with this setup is around 0.86 m. This enables a server node packet to reach a client node in fewer hops than that is achieved with low transmit power.

\subsection{Scenario 2: Vicinity range (0.6 m-0.86 m) with 0 dBm power}
The power level setting of each NRF52832  node is 0 dBm in this scenario. The vicinity range achieved with this setup is between 0.6 m-0.86 m. 
This results in a server node packet taking either 3 or 4 hops to reach a client node.

\subsection{Scenario 3: Vicinity range (0.6 m) with -4 dBm power }
In Scenario 3, we set the transmit power of NRF52832 server nodes to the least power level of -4 dBm. The vicinity range of the nodes is around 0.6 m. 
The diagonal nodes are 0.86 m away and hence lie outside the vicinity range of the transmit node. This results in a packet from a server node taking 4 hops to reach a client node.

\begin{figure}
     \centering
     \begin{subfigure}[b]{0.8\columnwidth}
         \centering
         \includegraphics[width=\columnwidth]{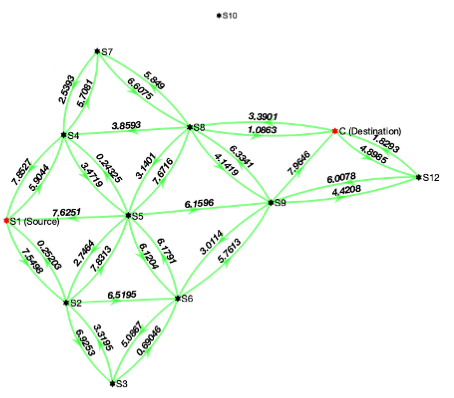}
         \caption{Node S10 separated}
         \label{fig:S10}
     \end{subfigure}
     \hfill
     \begin{subfigure}[b]{0.8\columnwidth}
         \centering
         \includegraphics[width=\columnwidth]{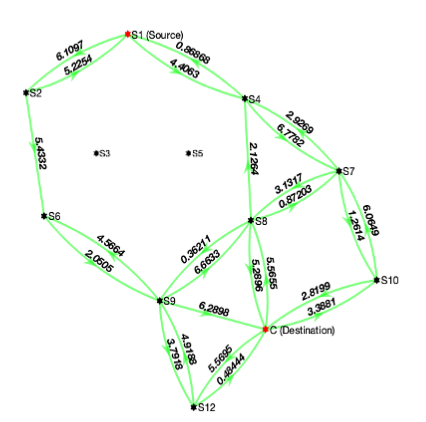}
         \caption{Node S3, S5 separated}
         \label{fig:S35}
     \end{subfigure}
     \hfill
     \begin{subfigure}[b]{0.8\columnwidth}
         \centering
         \includegraphics[width=\columnwidth]{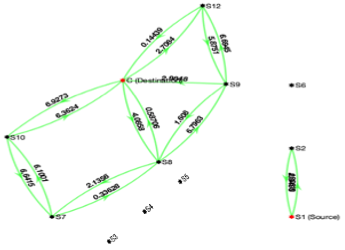}
         \caption{Node S3, S4, S5 and S6 separated}
         \label{fig:S456}
     \end{subfigure}
        \caption{De-tangled mesh network examples. Each de-tangled mesh shows separated nodes from the rest of the mesh network. For example, the second figure (b) shows the separation of nodes S3 and S5. The latency in each direction between the source-destination pair is shown along the edges for a single shortest-path iteration.}
        \label{fig:Detangled-mesh}
\end{figure}

\begin{algorithm}
\small
 \SetAlFnt{\small}
\SetAlgoLined
\caption{Compute PDR}
{
Input m \Comment{number of evidences for each i-hop network}\;
Input N \Comment{Maximum number of hops between source and destination}\;
Input $k_i$\Comment{fraction of i-hop network}\;
Input $PDR_i$\Comment{PDR of i-hop network between source and destination}\;
Input $L_s=[L_{1}, L_{2}, $\ldots$, L_{i}, $\ldots$, L_{N}]$ \Comment{Link set corresponding to each i-hop group between source and destination}\;
\While {($i \leq N)$}
{
\While {($j \leq m$)}
{
$b_j=BEL(C|S1)$\;
}
$PDR_i=\frac{sum(b_j)}{m}$\;
}
$PDR= PDR+(k_i \cdot PDR_i)$\;
}
\label{alg:anom_det_PDR}
\end{algorithm}

\begin{algorithm}[!h]
\small
 \SetAlFnt{\small}
\SetAlgoLined
 \caption{Network anomaly detection}
{
Input\quad $L_i (t_j)$ \hfill \Comment{latency of $i^{th}$ source node and the client node at the time period j}\;
Input\quad $PDR_i(t_j)$ \hfill \Comment{PDR of $i^{th}$ source node and the client node at the time period j}\;
Input\quad $So$\Comment{Source node list considered in a network}\;
Input\quad$N_i$\Comment{$i^{th}$ node in the network}\;
Input\quad $C$\Comment{client node}\;
Input\quad $b^{thl}_i$ \Comment{latency threshold of the $i^{th}$ node to the client node in a baseline mesh}\;
Input\quad $b^{thp}_i$ \Comment{PDR threshold of the $i^{th}$ node to the client node in a baseline mesh}\;
}
 \While{$i \in So$}{
 \uIf{$L_i(t_j) \geq b^{thl}_i$}{
  $Anom=1$\;
  } 
 \Else{
       \uIf{$PDR_i(t_j) \leq b^{thp}_i$}
        {
          $Anom=1$\;
         }
         \Else
         {
            $Anom=0$\;
            }
         }
  $i=i+1$\;
 }
\label{alg:anom_det}
\end{algorithm}
\section{De-tangled mesh detection}
\label{sec:detangled}
 A de-tangled mesh is a state of the mesh with one or more nodes being separated from the rest of the mesh network. This state can be detected with a decrease in reliability and an increase in the average latency between the client and server nodes in the mesh network. The Algorithm~\ref{alg:anom_det_PDR} and \ref{alg:anom_det} are used to compute PDR for each source-destination pair and identify a de-tangled mesh, respectively. {Fig.~\ref{fig:Detangled-mesh} shows three different de-tangled states for a 12-node mesh.
\subsection{Case 1: Belief-based reliability between the server node S1 and the client node C}
Consider Fig.~\ref{fig:S10}, which shows a de-tangled mesh network with the node S10 separated from the rest of the mesh network. 
We create an equivalent Bayesian network for a de-tangled mesh and compare it with the baseline mesh. To begin with, we consider node S1 as the source node in the mesh, and all other nodes between the server node S1 and client node C, act as relay nodes. We query the state of C with evidence $S1=1$ indicates that the source node S1 has transmitted a packet. 
A packet sent from S1 will take 0 to 136 paths to reach C in the considered baseline mesh network shown in Fig.~\ref{fig:int_add}. Here, each mesh link experiences different link failure probabilities. The number of paths between S1 to C is estimated using a Noisy-Integer addition function at each relay. Table  \ref{tab:rel_mesh_belief1} shows that separation of S10 affects the reliability of the mesh network, which is quantified with conditional probability of C taking a particular state s,  given the evidence S1, $BEL(C=s)=P((C=s)|(S1=1))$.
\begin{table}[h!]
    \centering
    \begin{tabular}{|c|c|c|}
    \hline
    Belief&Baseline& De-tangled\\
     for C's state &mesh&mesh \\
       \hline
        BEL(C=0) & 0.0203  & 0.0273 \\
        \hline
        BEL(C=135)&0.0243&0.0319\\
        \hline
        BEL(C=136)&0.0346&---\\
        \hline
    \end{tabular}
    \caption{Belief for different states of C representing the number of paths to reach C from the source node S1. Node S10 is separated in the de-tangled mesh. We compare beliefs between a baseline and a de-tangled mesh.  We infer from the first row that belief for no path to exist, BEL(C=0) increases from 0.0203 to 0.0273 after the separation of node S10. The second row shows the belief that 135 paths exist, BEL(C=135) is 0.0319 in a de-tangled network. The third row shows that the belief for C taking 136 paths, BEL(C=136) is 0.0346 in a baseline mesh network.}
    \label{tab:rel_mesh_belief1}
\end{table}

\subsection{Case 2: Belief-based reliability between the server node S7 and the client node C}
\begin{table}[h!]
    \centering
    \begin{tabular}{|c|c|c|}
    \hline
    Belief&Baseline& De-tangled\\
     for C's state &mesh&mesh \\
       \hline
        BEL(C=0) & 0.0067  & 0.0489 \\
        \hline
        \hline
        BEL(C=135)&0.0292&0.0177\\
        \hline
        BEL(C=136)&0.0183&---\\
        \hline
    \end{tabular}
    \caption{Belief for different states of C representing the number of paths to reach C from the source node S7. The first row shows that the belief for no path to exist, BEL(C=0), increases from 0.0067 to 0.0489 after the separation of node S10. The second row shows that the belief for 135 paths to existing BEL(C=135) is 0.0177 in a de-tangled mesh. We notice from the third row that the belief for C taking 136 paths, BEL(C=136), is 0.0183 in a baseline mesh network.} 
    \label{tab:rel_mesh_belief2}
\end{table}
 We consider S7 as the source node, and all other nodes between S7 and client node C act as relay nodes. We query the state of C with the evidence $S7=1$ indicating that S7 has transmitted a packet. S7 is closer to the separated node  S10 and experiences a larger reliability impact than the source node S1.

\begin {table}[h!]
    \centering
    \begin{tabular}{|c|c|c|c|c|}
    \hline
     & Average& Average&Average&Average\\
     & latency& latency& latency&latency\\
     Server& between  & between&between&between\\
     node& server & server &server & server\\
     number& and client & and client& and client& and client\\
    &in baseline & with S10&with S3 & with S3, S4,\\
          &&&and S5&S5 and S6\\
          & mesh & separated& separated& separated\\
          &(ms) & (ms) &(ms) & (ms)\\
    \hline
    S1& 11.16& 10.67&12.24&$\infty$\\
    S2& 9.15&8.32&10.96&$\infty$\\
    S3&10.04&9.76&--&--\\
    S4&8.79&8.66&9.93&--\\
    S5&7.28&5.77&--&--\\
    S6&6.85&6.56&7.38&--\\
    S7&6.13&7.28&6.10&5.98\\
    S8&3.97&3.75&3.56&3.56\\
    S9&6.04&3.40&3.44&3.47\\
       &6.08&&&\\ 
    S10&3.93&--&3.99&3.90\\
    S11& 4.02& 3.52&3.53&3.76\\
    \hline
    \end{tabular}
    \caption{The average latency between each source and the client node C is shown for a baseline and a de-tangled mesh. The latency between the server node S7 to client C has a larger difference in latency when compared to that of other server nodes. This indicates that S10 is a critical relay node for this server node in a baseline mesh. The overall latency variation from baseline mesh indicates the change in mesh configuration and the presence of an anomaly in  a mesh.}
    \label{tab:lat_server}
\end{table}

\subsection{Latency analysis for a de-tangled mesh}
Separation of a relay node from the mesh changes the latency between the source and the destination node. First, we characterize the latency for a de-tangled mesh and compare it with that of the baseline mesh. Subsequently, we monitor the mesh for a de-tangled state that involves the detection of change in latency.  We have considered three de-tangled cases: a)~S10, b)~S3 and S5, c) S3, S4, S5, and S6 nodes being separated from the rest of the mesh network.  
For each de-tangled case, the latency within the mesh is averaged over 1000 independent link uncertainty cases. Table~\ref{tab:lat_server} shows that the change in latency for the de-tangled mesh cases b)~and  c), is particularly high between server nodes S1, S2, and client node C.
Table~\ref{tab:det_latency} shows the overall average latency in the mesh evaluated for three different scenarios.\\ 
 \begin{table}[h!]
    \centering
    \begin{tabular}{|c|c|c|}
    \hline
     Vicinity & Average latency& Average latency \\
     range& in baseline mesh & in de-tangled mesh\\
     (m) & (ms) & (ms)\\
    \hline
       Scenario 1 & 6.25& 6.46 \\
    \hline
       Scenario 2 & 6.96 &7.43 \\
    \hline
       Scenario 3 &8.07 &8.33 \\
    \hline
    \end{tabular}
    \caption{BLE mesh latency in a baseline and de-tangled state is shown for various vicinity ranges 0.86 m (Scenario 1), 0.6 m -0.86 m (Scenario 2), and 0.6 m (Scenario 3). The simulations incorporate vicinity ranges and the associated mesh connections. For example, Scenario 1, 2, and 3 vicinity ranges translate to no connection, intermittent connection, and stable connection with diagonal nodes, respectively. The adjacent nodes are always connected in all three scenarios.}
    \label{tab:det_latency}
\end{table}

\section{PLC network characterization}
\label{sec:PLC_char}
The PLC network is the backbone that connects several spatially isolated dense meshes. We now characterize the PLC network using a measurement-aided channel frequency response model.
\subsection{PLC channel frequency response}
\label{sec:PLC_cfr}
\begin{figure}
    \centering
    \includegraphics[width=0.8\columnwidth]{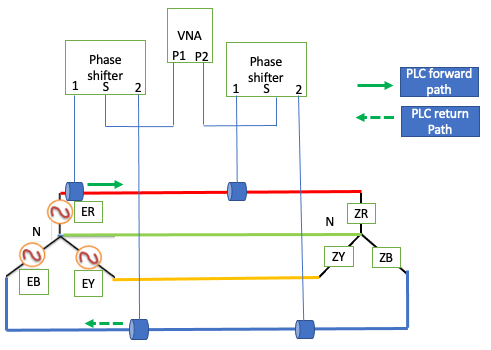}
    \caption{PLC network setup with the 3-phase power supply connected to symmetrical 4-core cable loads $Z_R$, $Z_Y$, and $Z_B$. A VNA is used to measure two-port S-parameters of the network.}
    \label{fig:plc_network}
\end{figure}
We perform a two-step procedure to create a large synthetic Channel Frequency Response (CFR) data set.
\begin{itemize}
    \item S-parameter measurements in a cable section with various loads (small dataset from measurements).
    \item Synthetic CFR data obtained from S-parameters after considering loads in all branches of the cable.
\end{itemize}
We perform an online measurement campaign on a cable section with various loads that switch 
on and off in random intervals. Our experimental PLC setup includes a symmetrical four-core cable excited using 415~V 50~Hz three-phase power supply with 0.2 and 0.1~kW loads. Fig.~\ref{fig:plc_network} shows the PLC network setup for S-parameter measurements. 
We provide  the detailed method to generate synthetic CFR from a small measurement dataset \cite{huo2021measurement}. We generate synthetic CFR data from S-parameters and use that to derive PDR and latency measures. These measures characterize the PLC network.

\subsection{PLC latency model}
\label{sec:plc_lat}
We compute the root mean squared delay spread $\tau_{rms}$ of the powerline derived from the CFR \cite{tonello2014home}. Since $\tau_{rms}$ is a function of the power delay profile p($\tau$), we first compute the CIR $h(\tau)$, defined as N-Point IFFT of CFR $H(f)$. In our work, we consider a 40 MHz spectrum with a frequency resolution $\delta f$ 200 kHz, resulting in a time resolution T, 5 microseconds.
\begin{equation}
\label{eq:rms_delay_spread}
\begin{aligned}
    h(\tau)&=IFFT(H(f)) \\
    \tau&=mT, m=0,1,2, \ldots , N\\
\end{aligned}
\end{equation}
\begin{equation}
    p(\tau)=\frac{|h(\tau)|^2}{\sum_{m=0}^{N} |h(\tau)|^2}
\end{equation}
\begin{equation}
\begin{aligned}
    E(\tau^2)&= \sum_{m=0}^{N} (\tau)^2 \cdot (p(\tau))\\
     E(\tau) &= \sum_{m=0}^{N} (\tau) \cdot p(\tau)\\
    \tau_{rms}&=\sqrt{(E(\tau^2)-(E(\tau))^2}\\
    \end{aligned}
\end{equation} 
The maximum likelihood estimate of our experimental data is well represented using Lognormal distribution.
\subsection{PLC reliability model}
\label{sec:plc_rel}
The PLC reliability model is formulated as a function of the Signal to Interference Noise Ratio (SINR) in the powerline. SINR can be represented as a function of transmitted power, received power, Interference power, and Noise floor at the receiver, and the average channel gain.\\
 $G$: Average channel gain\\
 $P_s$: Transmitted signal power\\
 $P_n$: Noise power at the receiver\\
 $I$: Interference power\\
 \begin{equation}
     SINR_{plc}= G\frac{P_s}{I+P_n}
 \end{equation}
 where the average channel gain is a function of Channel Frequency Response (CFR), $H(f)$, $f$ is the frequency point in the spectrum as indicated in \cite{tonello2014home}.\\
 \begin{equation}
    G=\frac{1}{N} \sum_{f=1}^{N} |H(f)|^2 
 \end{equation}
 where $N$ is the number of frequency points considered in the spectrum.
  G is computed as the square of the magnitude of CFR and its summation over the number of frequency points.
  Our experimental results show a gain of the channel as -34.36 dB, 
 The transmit power of -55~dBm/~Hz translates to $3.162 \times 10^{-12}$~Watts/Hz, and the noise power over 2-40 MHz corresponds to 0.13 Watts/Hz.
 Eq.~\ref{eq:noise_power} shows the frequency-dependent noise power at the receiver. 
 \begin{equation}
 \label{eq:noise_power}
     P_n(f)= 10 log_{10} (\frac{1}{(m \delta f)^2 \times 10^{-15.5}}) \quad dBm/Hz
 \end{equation}
 The Bit Error Rate (BER) is a function of $SINR_{plc}$.
 \begin{equation}
 BER= Q(2\sqrt{SINR_{plc}})
 \end{equation}
 Where Q(.) is the complementary error function.
 \begin{equation}
      PDR_{plc}= 1-(BER)^N
 \label{eq:pdr_plc}
 \end{equation}
 Where N is the number of bits in a packet, we consider PLC transmission, with each packet having 136 bits and is transmitted using BPSK communication.
The PDR is computed using BER under the assumption of Independent and Identical Distribution (i.i.d).

 \section{Hybrid network reliability}
 \label{sec:hybrid_rel}
 Let the mesh network coverage area required for an application be $l \times b$. We then apply a grid by dividing each $l\times b$ into $l_1\times b_1$ with length of the grid, $l_1=\frac{l}{m_1}$ and breadth of the grid $b_1=\frac{b}{m_2}$ where $m_1$ and $m_2$ are integers. $Grid_{size}=l_1\times b_1$ is the area of the reference mesh grid. Hence, the number of mesh units, $\alpha$, that spans the network coverage area depends on the $Mesh_{area}$ and $Grid_{size}$. $\beta$ is the number of PLC hops to cover the distance  required in addition to the mesh network. Fig.~\ref{fig:mesh_config_unit_span} shows the mesh configuration divided into many mesh units.
 \begin{figure}
     \centering
     \includegraphics[width= \columnwidth]{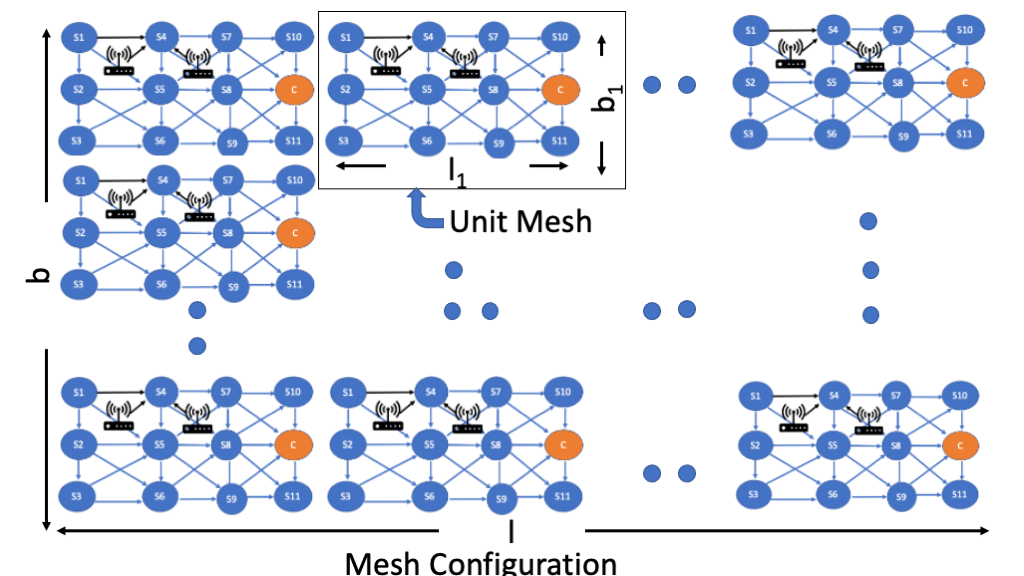}
     \caption{BLE mesh with grid size $l_1 \times b_1$ is used to span the entire network coverage area of size $l \times b$. The PDR averaged over 1000 independent RF interference scenarios is considered for numerical simulation.}
     \label{fig:mesh_config_unit_span}
 \end{figure}
 \begin{equation}
\label{eq:hybrid_pdr}
\begin{aligned}
    PDR_{hybrid}&= (PDR_{ble})^\alpha (PDR_{plc})^\beta\\
\end{aligned}  
\end{equation}
We use Eq.~\eqref{eq:mesh_pdr} to evaluate the $PDR_{ble}$ within a mesh unit.  We use Eq.~\eqref{eq:pdr_plc} and evaluate the $PDR_{plc}$ of the PLC network.  The number of PLC hops and number of mesh units are used to evaluate the end-to-end $PDR_{hybrid}$ of the network.
\begin{equation}
\begin{aligned}
 PDR_{hybrid}&=(\sum_{i=0}^{N}k_i \cdot PDR_i)^\alpha  (1-(BER_{plc})^n)^\beta \\
    \alpha&= Mesh_{area}/ Grid_{size}\\
\end{aligned}
\end{equation}

N: number of hops between client and server nodes.\\
$PDR_i$: The average PDR of the i-hop path groups between mesh client and server nodes.\\
$\alpha$: number of mesh units.\\
$\beta$: number of hops in the PLC network. \\
n: number of bits in a PLC packet.\\
\begin{table}[htbp]
    \centering
    \begin{tabular}{|c|c|c|c|c|c|c|}
    \hline
       Vicinity&$\alpha$& Mesh&$PDR_{ble}$&$\beta$&$PDR_{plc}$&$PDR_{hybrid}$\\
       range&&&&&&$\{\alpha = 1$\\
       &&length&&&&, $\beta\}$\\
     &&(m)&(\%)&&(\%) & (\%)\\
       \hline
        &1&2.4&97.38&1&& 91.3\\
       &2&4.8&94.8&2&93.8&85.7\\
    Scenario 1&3&7.2&92.3&3&&80.4\\
       &10&24&76.7&&&\\  
       \hline
       &1&2.4&96.95&1&&91\\
    Scenario 2&2&4.8&94&2&93.8&85.2\\
     &3&7.2&91.1&3&&80\\
     &10&24&72.9&&&\\
      \hline
       &1&2.4&93.2&1&& 87.4\\
    Scenario 3&2&4.8&86.8&2&93.8&82\\
       &3&7.2&80.9&3&&76.9\\
       &10&24&49.44&&&\\
     \hline
    \end{tabular}
    \caption{PDR of the hybrid network with various vicinity ranges of the mesh node, number of mesh units ($\alpha$), and the number of plc hops ($\beta$). $PDR_{hybrid}$ is calculated using the Eq.~\eqref{eq:hybrid_pdr}. The scaling of the mesh by three times in Scenario 1 results in ~5\% reduction in $PDR_{ble}$. However, scaling of the mesh by three units in Scenario 3 leads to ~11\% reduction in $PDR_{ble}$, and scaling of the mesh by ten units results in a drastic reduction of $PDR_{ble}$ across all scenarios. Alternatively, if one mesh unit combined with a PLC hop, achieves long range with a $PDR_{hybrid}$ of around 91\% in Scenarios 1 and 2.}
    \label{tab:hybrid_pdr}
\end{table}

\section{Hybrid network latency}
\label{sec:hybrid_lat}
 Latency in the hybrid network to detect a de-tangled state of an independent mesh, is represented as a sum of latency in the independent mesh, the latency associated with the gateway, and the latency in PLC network. Each independent mesh can have several mesh units ($\alpha$).  Fig.~\ref{fig:lpwan_ch5} shows a BLE mesh network and several PLC hops based on the distance to be covered. For example, different cargo zones to CMS in air cargo monitoring application.

\begin{equation}
\begin{aligned}
   L_{hybrid}&={\alpha}L_{ble}+ L_{gw} +\beta L_{plc}\\
                &={\alpha}L_{ble} + L_{gw} +\beta(\tau_{rms}+L_{relay})\\
                \alpha&= Mesh_{area}/ Grid_{size}\\
                \beta&=(plc_{nodes}-1)\\
\end{aligned}
\end{equation}
where $L_{ble}$ is the latency evaluated in a reference mesh unit. Let the  reference mesh cover an area, $Grid_{size}$. For example, in our case $Grid_{size}=2.4 m \times 1.8 m$. The $Grid_{size}$ is scaled in integer multiples to span the network coverage area. $L_{gw}$ is the latency associated with the gateway for BLE to PLC protocol conversion and vice versa. $L_{plc}$ is the latency associated with the PLC link and is represented as multiples of latency in the reference link. We have considered a 100 m link as a reference link. $L_{relay}$ is the latency associated with the relay node. $\beta$ represents the number of PLC hops. If the PLC link has switching loads, then $\tau_{rms}=10.8$ ms is the associated latency. While the PLC link has resistive loads, then $\tau_{rms}=6$ ms. 
\begin{table}[htbp]
    \centering
    \begin{tabular}{|c|c|c|c|c|c|c|}
    \hline
      Mesh& PLC& $\alpha$&$L_{ble}$ & $\beta$&$L_{plc}$&$L_{hybrid}$\\
       &&&&&&$\{\alpha=1,\beta\}$\\
       states &loads    &  & (ms) & &  (ms) & (ms)\\
        \hline
       & &1 &8.4  & 1 & 1.78  & 10.15 \\
      Dense&Resistive&2&16.8&2&3.77&12.17\\
      &&3&25.2&3&5.34&13.74 \\ 
      &&10& 84.0& 1& 1.78& 10.15\\\cline{2-7}
      &Switching&1&8.4&1&5.66& 14.07\\ 
      &&2&16.8&2&11.57&  19.97 \\
      &&3&40.5&3&16.98&25.38\\
      &&10&84.0&1&5.66&14.07\\
      \hline
      &&1& 13.5&1&1.78& 15.28\\ 
      &Resistive&2&27.0&2&3.77&17.27\\  
    &&3&40.5&2&5.34& 18.84\\ 
      &&10 &135.0&1&1.78& 15.28\\ \cline{2-7} 
       Sparse   &Switching& 1&13.5&1&5.66&19.32\\
     & &2&27& 2& 11.57&25.07\\
     &&3&40.5&3&17.25& 30.76\\
        &&10&135&1&5.66&19.32\\
      \hline
    \end{tabular}
    \caption{Numerical simulations using end-to-end hybrid network latency model that considers the number of mesh units ($\alpha$) and the number of plc hops ($\beta$). The scaling of a dense mesh by three units leads to a network coverage length of 24~m with mesh latency, $L_{ble}$ of 84 ms . An alternative method for scaling mesh with a PLC network will yield an 8$\times$ reduction  in latency, $L_{hybrid}$.}
    \label{tab:hyb_lat}
\end{table}

\section{Discussions and insights}
\label{sec:disc}
The primary goal in this paper is the detection of a de-tangled mesh. 
 When compared to the characterized baseline mesh, a significant deviation in the PDR and latency measures indicates a de-tangled state of the mesh.
 We answer several research questions through our extensive simulation study using causal inference on an apriori network based on a real-world dataset collected from a cargo monitoring application \cite{mala2019heterogeneous}. \\
 \begin{itemize}
     \setlength{\itemsep}{10pt}
     \item RQ1: \textit{How does the performance change to indicate a de-tangled state in a mesh?}
     \item RQ2: \textit{What is the severity of a de-tangled mesh state?}
     \item RQ3: \textit{What is the process to localize a de-tangled mesh state?}
     \item RQ4: \textit{How to deal with scalability challenges in a mesh?}    \\
 \end{itemize}
\noindent \textbf{RQ1: De-tanglement Identification} \\
Table~\ref{tab:rel_mesh_belief1} shows the belief-based reliability in a de-tangled mesh, BEL(C=s) with evidence S1, and Table~\ref{tab:rel_mesh_belief2} shows the BEL(C=s) with evidence S7. While observing the first row of these two tables, we infer that the belief for zero or no path between the source and client node increases for a de-tangled mesh. Furthermore, since the de-tangled  mesh node S10 is between S7-C,  the evidence $S7=1$ has  seven times more belief for no path. Increase in belief for no path indicates separation of nodes in the mesh. \\
\noindent \textbf{RQ2: Severity analysis} \\
We infer from Table~\ref{tab:lat_server} that when more number nodes are separated from the mesh, the latency difference from the baseline mesh increases. For example, the average latency between nodes S1-C in the baseline mesh is 11.16~ms. However, the average latency of the same source-destination pair in a de-tangled mesh with (a) S10, (b) S3 and S5, (c) S3, S4, S5, and S6 nodes being separated are 10.67~ms, 12.24~ms, and $\infty$ respectively. Here, the severity of the de-tanglement can be inferred from the latency. The case (c) S3, S4, S5, and S6 separated nodes have a more severe impact than case (a) S10 being separated. Based on severity, countermeasures can be suggested.\\ 
\noindent \textbf{RQ3: Localization analysis}\\ 
Similarly, we infer from Table~\ref{tab:lat_server} that when the separated nodes are between the source and the destination, then it increases the latency. Otherwise, the latency remains constant. For example, the latency between server node S7 and the client in baseline mesh is 6.13 ms, while the latency in a de-tangled mesh (a) S10, (b) S3 and S5, (c) S3, S4, S5, and S6 nodes being separated are 7.28, 6.10, and 5.98 ms respectively.  The node S10, being a relay node between S7-C, increases the latency while S3, S4, S5, and S6 are not in the path of S7-C, and hence the latency corresponding to those de-tangled mesh cases are similar to that of the baseline mesh.  Hence, these latency measure helps us to localize the de-tangled mesh part in a large MoT network. \\
If a critical node for a source node is de-tangled, then the latency between a source and a client node varies significantly. Otherwise, the latency is not impacted. Hence, mapping of critical nodes and source-destination pair will eventually help in the localization of de-tangled mesh. For example, S10 is the critical node for the S7-C source-destination pair, and hence the de-tanglement of this node impacts the latency significantly. De-tanglement of nodes S3, S4, S5, and S6 does not affect the latency, indicating that these nodes are not critical nodes for the source-destination pair. If one wishes to identify the de-tanglement of these nodes, node S1 and S2 should be chosen as source nodes with a C as the destination node. \\
\noindent \textbf{RQ4: Scalability analysis}\\
Table~\ref{tab:det_latency} shows a change in average latency between a baseline and a de-tangled mesh. In this case, the source node S1 communicates to the client node in a mesh where each node has the same transmit power that translates to the corresponding vicinity ranges from Scenario 1-3. 
Table~\ref{tab:hybrid_pdr} shows the PDR of mesh, PLC link, and hybrid MoT network. The scaling of the mesh by three units in Scenario 1 results in ~5\% reduction in PDR. However, in Scenario 3, the three units scaling of mesh  results in ~11\% reduction in PDR. The 10-unit mesh achieves 24~m network coverage and results in a drastic reduction in PDR in all scenarios. Instead of a large mesh, a single mesh unit with a PLC hop extends the range of the mesh with better $PDR_{hybrid}$ than $PDR_{ble}$. We infer from Table~\ref{tab:hyb_lat} that when the mesh scales by ten units, the latency increases to 84 ms. A method to scale to the same distance with a hybrid link results in an 8$\times$ reduction in latency and a PDR of 91.3\%. Hence we propose to generate a large mesh coverage area by interconnecting several smaller meshes using a PLC network. This solution results in comparatively better $PDR_{hybrid}$ and $L_{hybrid}$ than $PDR_{ble}$ and $L_{ble}$.

\section{Conclusions}
In this paper, we present the PDR and latency models to evaluate the performance of the Mesh of Things (MoT) network through numerical simulations. First, we characterize a baseline mesh with these performance measures. Subsequently, the continuous tracking of these measures facilitates the identification of a de-tangled mesh. Both the movement of nodes away from the rest of the mesh network and RF interference will affect the latency and PDR measures of the mesh. 
We propose the PDR and latency model for a hybrid MoT network. This network is a good candidate for implementing a low-power wide area network to track performance metrics and reliably detect network anomalies.
\section*{Acknowledgements}
The authors thank Dr. Chandramani Singh for his valuable comments.
\ifCLASSOPTIONcaptionsoff
  \newpage
\fi
\bibliographystyle{IEEEtran}
\bibliography{IEEEabrv,Bibliography}
\end{document}